\documentclass[conference]{IEEEtran}
\IEEEoverridecommandlockouts
\usepackage{cite}
\usepackage{amsmath,amssymb,amsfonts}
\usepackage{algorithmic}
\usepackage{graphicx}
\usepackage{textcomp}
\usepackage{xcolor}
\usepackage{comment}
\usepackage{hyperref}
\usepackage[hang,flushmargin]{footmisc}
\def\BibTeX{{\rm B\kern-.05em{\sc i\kern-.025em b}\kern-.08em
    T\kern-.1667em\lower.7ex\hbox{E}\kern-.125emX}}
\begin{document}

\newcommand\todo[1]{\textcolor{red}{#1}}

\title{A Recommender System For Open Educational Videos Based On Skill Requirements}

\author{
    \IEEEauthorblockN{Mohammadreza Tavakoli, Sherzod Hakimov, Ralph Ewerth, G\'abor Kismih\'ok}
    \IEEEauthorblockA{Technische Informationsbibliothek, Germany \{reza.tavakoli, sherzod.hakimov, ralph.ewerth, gabor.kismihok\}@tib.eu
    }
}

\maketitle

\begin{abstract}
In this paper, we suggest a novel method to help learners find relevant open educational videos to master skills demanded on the labour market. We have built a prototype, which 1) applies text classification and text mining methods on job vacancy announcements to match jobs and their required skills; 2) predicts the quality of videos; and 3) creates an open educational video recommender system to suggest personalized learning content to learners.

For the first evaluation of this prototype we focused on the area of \emph{data science} related jobs. Our prototype was evaluated by in-depth, semi-structured interviews. 15 subject matter experts provided feedback to assess how our recommender prototype performs in terms of its objectives, logic, and contribution to learning. More than 250 videos were recommended, and 82.8\% of these recommendations were treated as useful by the interviewees. Moreover, interviews revealed that our personalized video recommender system, has the potential to improve the learning experience.
\end{abstract}


\section{Introduction}

In the recent decades, we have been facing an indisputable change in the quantity and quality of skills demand and supply\cite{colombo2018applying}. This dramatic change comes with a number of educational challenges for all labour market stakeholders (e.g. educational decision makers, employers, and employees). The gap, for instance, between skill demands and educational programs \cite{wowczko2015skills,mcgill2009defining} is eminent. Moreover, the existence of transparent information on skill requirements should be a deep concern of our current societies and citizens alike. This information has significant importance to those, who want to secure long term employability, achieve promotions in their current workplace \cite{colombo2018applying,khobreh2015ontology}, or, in the light of the current COVID-19 pandemic, to those, who need to re-skill themselves to adapt to post COVID-19 labour markets (and regain employment).
In order to supply stakeholders with labour market information, a number of occupational taxonomies (e.g. ESCO, O*NET) exists to provide information to job-holders and job-seekers about occupations. However, most of these taxonomies are manually maintained, therefore they are time-consuming, expensive and also susceptible to being outdated \cite{djumalieva2018open}.

At the same time, literature shows that open education is gaining traction in the area of personal skills development \cite{kanwar2018global}. Open Educational Resources (OERs), especially videos, as content sources have become popular for learners \cite{ha2011novel}. They are supplied by large amount of experts around the world, with different levels of expertise in a wide range of professional contexts (e.g. discipline, location or language). However, the uptake of OERs is limited in most user groups (e.g. educators or learners) \cite{sun2018heuristic,ruiz2014semantically,chicaiza2015user,ha2011novel,chicaiza2017recommendation}, as OER repositories still struggle to provide personalised services for their users. As an illustration, if a learner wants to find appropriate learning content, (s)he must search manually through several OER repositories with different interfaces. Furthermore, the few OER recommendation algorithms available are limited to approaches such as building (or reusing existing) ontologies \cite{wan2018learning,sun2017towards}, analysing user behavior in social networks \cite{lopez2014recommendation}, or applying text mining techniques to identify similar OER documents \cite{duffin2007oer}.

Most of the research on learning content recommendations focus on textual resources, similarly to the field of search as learning \cite{hoppe2018}. However, insights from educational psychology and multimedia learning \cite{mayer2005cognitive} suggest that images and videos might be preferable when addressing specific learning needs (e.g., procedural learning tasks \cite{genuchten2012}). For many topics a large amount of educational and lecture videos are available on the Web, ranging from single video tutorials to entire Massive Open Online Courses (MOOCs). However, when exploring the Web and video platforms it is difficult to find (high-quality) videos that are tailored to the user's (required) skills and knowledge levels.



In this paper, therefore, we address the above mentioned challenges and build a software prototype to provide personalized OER (specifically videos) recommendations that help learners to master skills needed for their current or future jobs. Thus, the main objectives of this paper are:

\begin{itemize}
    \item Empower learners to construct their own learning trajectories on the basis of labour market information and OERs
    \item Create an algorithm to decompose jobs into unique skills
    \item Build a model to predict the quality of video based OERs 
    \item Develop and evaluate a personalized open educational video recommender system prototype, relying on labour market information and videos' properties
\end{itemize}

This paper is organized as follows: Section \ref{related-work} discusses the state-of-the-art of OER recommender systems. Subsequently, in Sections \ref{data-collection} and \ref{method}, we explain the processes of data collection and the construction of our recommender system. Section \ref{validation} shares the validation results of our prototype. Finally, we conclude the paper, and define our future steps in Section \ref{conclusion}.


\section{Related Work}\label{related-work}
Based on available literature \cite{chicaiza2017recommendation} there is an enormous development potential in building OER-based recommender systems due to the vast amount of open educational content available globally and also due to their currently limited functionalities. Moreover, there is no signal that typical lifelong learning factors (skills, jobs) play any role in current work. We grouped the relevant available studies into the following three categories:
\subsubsection{Semantic and Ontology Based Methods}
Some studies make use of ontologies, linked data, and open source RDF data to leverage semantic content, and define recommendation algorithms \cite{chicaiza2017recommendation,ruiz2014semantically,sun2017towards,chicaiza2015user}. For instance, \cite{wan2018learning} builds an ontology for learners, learning objects, and their environments to establish similarity measures between learning objects, update their properties, and provide diverse and adaptive recommendations.

\subsubsection{Social Network Analysis}
\cite{lopez2014recommendation} builds graphs of OERs and learners based on social networks. They find tweets with valid educational URLs, and build an OER graph based on the co-occurrences of hashtags. They also use \emph{mentions} and \emph{retweets} to build a learner graph. Finally, they recognize important and influential nodes, and use density and centrality measures from the graph to provide recommendations.

\subsubsection{Machine Learning Based Methods}
 \cite{duffin2007oer} finds similar OERs using Document Clustering and LSA and makes recommendations based on the similarities. Moreover, \cite{cooper2018moocex} considers video content and sequential topic relationships to provide courses across multiple platforms while \cite{Zhu2013VideoTopicCV} uses a topic modeling algorithm to recommend videos based on user context and interests.

\section{Data Collection}\label{data-collection}
In this section we describe the data we collected to build an open educational platform to recommend educational videos. Firstly, we describe the procedure of collecting skills, followed by an explanation on the retrieval of educational videos. 

\subsection{Skill Collection}\label{skill-collection}
The first step for building our recommender was the identification of skills that are correlated with particular jobs. We used online job vacancies and built a model to extract skills for jobs dynamically and avoid any dependency on existing taxonomies (which are susceptible to outdating). To build such a model, we used a crawled sample dataset from Monster.com containing 22,000 job vacancies\footnote{\url{https://www.kaggle.com/PromptCloudHQ/us-jobs-on-monstercom}}. After an exploratory analysis, we concluded that large number of vacancies do not contain a "Required Skills" section. Therefore, we selected vacancies explicitly containing a "Required Skills" section and used them to build a classification model to detect sentences, which define skill requirements. To build the model, at first, we run the following pre-processing procedure: 

\begin{itemize}
    \item Removal of unimportant characters, punctuations, stop words
    \item Sentence tokenization, lowercase conversion and lemmatisation
\end{itemize}

Altogether we obtained more than 60,000 sentences including both sentences, which were mentioned in a "Required Skills" section (around 15,000 sentences and we set their label to 1), and also sentences mentioned in other sections in vacancies (around 45,000 and we set their label to 0). 

We trained a binary classifier model using FastText library in Python for our classification task \cite{joulin2016bag}. The classifier uses word n-grams and learns embeddings as a training process. The obtained dataset was split as 80\% for training and 20\% for evaluation. Applying our model to the test dataset showed that our model can detect skill-related sentences with F1 score of \textbf{88.7\%} (harmonic mean of precision and recall). 

Consequently we used the trained classifier to detect sentences that contain skill related content. After identifying those sentences, we applied TF-IDF weighting (with \emph{Minimum Document Frequency} of 3 as cut-off point) to detect skill terms in skill related sentences. All n-grams from the classified sentences were scored and the highest ranking n-grams are extracted as skill terms.



We ran our skill extraction method on 300 randomly crawled data science job vacancies (the context of our first prototype), which have been published on Monster.com in December 2019 and obtained a list of skills that learners should focus on for building a career in data science. In total, we extracted 16 important and unique data science skills. We provide a sample skill with other metadata below.

\begin{itemize}
    \item[] \textbf{Skill}: Python programming
    \item[] \textbf{Keywords}: python, python programming
    \item[] \textbf{Description}: Python is an interpreted, high-level, general-purpose programming language.
\end{itemize}

To find skill descriptions we used Wikipedia python API\footnote{\url{https://pypi.org/project/wikipedia/}} and crawled the wikipedia content, which are related to skills\footnote{Complete list of skills with their properties is available: \url{https://github.com/rezatavakoli/ICALT2020_recommender}}.

\subsection{Open Educational Videos}\label{video-collection}
 We collected educational videos from two main sources: YouTube and TIB AV portal\footnote{For this prototyping exercise, we used openly available videos, but we disregarded the type of license for our analysis. Nevertheless, licensing will obviously play a role in future implementations.}. YouTube is the most popular platform for hosting any type of video content. The TIB AV-Portal\footnote{\url{https://av.tib.eu/}} is a dedicated portal for scientific videos from the realms of architecture, chemistry, computer science, engineering and technology, mathematics, etc. and the videos include among others, computer visualisations, learning material, simulations, experiments, interviews, video abstracts, and recordings of lectures and conferences.

We retrieved videos by performing a keyword search on each portal. As explained above, each skill contains a set of keywords. All keywords were used to search and retrieve relevant videos. Videos from both sources might contain transcriptions of audio. YouTube includes them as subtitles, TIB AV Portal shows the body of the transcribed text. Upon availability, we extracted these transcriptions for retrieved videos. Missing transcriptions were obtained by applying Google Cloud Speech\footnote{\url{https://pypi.org/project/google-cloud-speech/}} on audio files extracted from videos.

Videos contain different types of information depending on the source. We collected/calculated the following metadata from YouTube and TIB AV portal videos.
\begin{itemize}
    \item[] \textbf{YouTube}: title, target skill, URL, length, description, transcription, view count, rating, likes, dislikes, relevancy score (assigned according to the rank in the search results), Textual similarity (which is calculated based on the similarity between skill description and video transcription. We explain this calculation in Section \ref{method})
    \item[] \textbf{TIB AV Portal}: title, target skill, URI, description, transcription
\end{itemize}

For our first prototype, we retrieved 550 videos from YouTube and 57 videos from TIB AV portal, which mentioned the 16 skills we fetched previously. These videos were presented to six experts in data science (with more than six years of industrial and more than three years of teaching experience in data science related positions) to annotate whether they fit to their target skill, or not. Each video was reviewed by at least three annotators and annotators assigned at least 2 minutes to set the label of each video. The final label was assigned based on a majority vote. In total, the annotators provided labels for 550 videos, where 213 of them fit to a skill (positive label) and 337 do not fit (negative label). The complete list of videos and labels are available for the research community\footnote{Annotated dataset including videos' properties and labels: \url{https://github.com/rezatavakoli/ICALT2020_recommender}}.

\section{Method}\label{method}
The following section provides details about how the recommender system was built, and how learners interact with our learning dashboard.

\subsection{Fit Prediction}\label{quality-prediction}
We trained a machine learning model to predict whether a given video fits to a skill or not. As mentioned earlier, we selected 550 videos for 16 skills to annotate whether a video fits a skill or not and annotators provided labels for 213 videos as a fit and 337 were annotated as not fitting. 

A Random Forest model was trained on the annotated data to build a model that outputs a binary decision: match/no-match. The algorithm uses the following video features to train our model.

\begin{itemize}
    \item Length: the length of a video in seconds
    \item Rating: the user rating, what a video received on a platform
    \item View count: the number of views on a video
    \item Relevancy score: the score assigned during the search process based on the video-platforms' results ranking as $\frac{1}{ranking\_position}$
    \item Level: the pre-defined levels either \textit{beginner}, \textit{intermediate} or \textit{advanced}. The levels are set during collection process by concatenating the search term with ``beginner``, ``intermediate`` or ``advanced`` to search for videos at different levels.
    \item Text similarity: the similarity is computed between skill description and video transcription. First, each word in a text is encoded using pre-trained 300-dimensional Glove vectors \cite{pennington2014glove}. Second, we average the vectors of words in a text to get a single vector that represents the whole text. We apply the described method to obtain a vector representation for both video transcription and skill description. Finally, the text similarity is a cosine similarity between the resulting two vectors.
\end{itemize}

70\% of the data was used to train the model and the remaining 30\% was used for the evaluation. The classifier achieves F1 score of \textbf{86.3\%} in predicting whether a video matches a skill. Additionally, we analysed 
the importance of each feature for the classification task. The trained model assigned different importance score to each feature on the basis of the provided training data. Each feature has a different weight on the decision based on these scores. The weights are calculated by pruning out trees below a particular node (as feature selection). The weights for the selected features are calculated as follows: length: 0.61, rating: 0.10, view count: 0.10, relevancy score: 0.08, level: 0.2, text similarity: 0.09. The model assigns the highest score for the \textit{Length} feature and it is followed by the \textit{Rating} feature. In the following section we describe how the trained binary classifier can be used within a recommender system.

\subsection{Recommendation Generation}\label{recommender-system}
Our proposed recommender system suggests new content to learners based on different parameters. The goal is to optimize weights for these parameters by increasing learner satisfaction (based on their ratings). The recommender system uses the following parameters:
\begin{itemize}
    \item Popularity: We calculate the difference between the number of likes and the number of dislikes for each video, group videos by their target skill, and for each group we normalize the values using Minmax normalization
    \item Fit probability: The probability of a video fitting a skill (explained in Subsection \ref{quality-prediction})
    \item Length: the length of the videos; we group videos by their target skill, and for each group, we normalize the length using Minmax normalization
    \item Text similarity: the textual similarity between a video transcription and a skill description
\end{itemize}

We build a 4-dimensional vector of $X$ where each item in the vector is value for a parameter mentioned above. We define a vector $P$ as a preference matrix for each user that contains a weight for each parameter in $X$. The goal is to optimize weights in $P$ for each learner based on previous ratings. In this way, we capture learners' preferences to provide personalised recommendations. The following loss function is used to optimise the weights in $P$ for each user with respect to the generated recommendations.

\begin{equation} \label{eq:loss}
  Loss Function = \sum_{i=recommendations} |(P * X_i) - Y_i|
\end{equation}
where $X_i$ is the mentioned 4-dimensional vector of a recommended video \emph{i} and $Y_i$ is the satisfaction rate of the user for that particular video \emph{i}. We use Gradient Descend to find the best $P$ for each user and the initial weights in $P$ are set by taking weights from similar users (e.g. with the same job, location, etc.).

Finally, the system generates a recommendation based on optimised weights in the preference matrix $P$ given the parameters of each video for a target skill. The videos are ranked by computing a cosine similarity between their matrix $X$ and the preference matrix $P$ of a user. The video with the highest score is given as a recommendation to a user.

\subsection{Learning Dashboard}\label{recommendation-serving}
We have built the prototype of our recommender system in the form of a dashboard\footnote{Demo of our prototype is available: \url{https://github.com/rezatavakoli/ICALT2020_recommender}}. The users interact with the dashboard for searching or adding skills they want to master, setting their levels of expertise for each skill, and adding contextual information about their occupation, geographical location and educational level. Subsequently, on the learning tab, the dashboard shows the list of their target skills, the learner's current expertise levels in each of them, and the links to the recommended open educational videos. Learners can watch the recommended content or ask for a new one in case they are not satisfied with the recommendations.

\begin{figure}
    \centering
    \includegraphics[height=23mm]{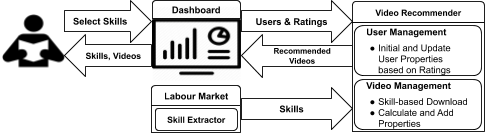}
    \caption{Components of our Labour Market based Video recommender}
    \label{fig:structure}
\end{figure}

After watching a recommended video, the learner rates her satisfaction with the recommendation. The system changes the learner's expertise level, updates her preference matrix $P$, and provides a new recommendation based on the new expertise level and preferences. The process continues until the learner reaches the highest mastery level for a particular target skill. Figure \ref{fig:structure} depicts the building blocks of our proposed approach.
\section{Validation}\label{validation}
To validate our proposed approach, we conducted semi-structured interviews with subject matter experts in the job area of \emph{Data Science}. We randomly selected 300 job vacancies related to data science from Monster.com which have been published in December 2019. Afterwards, we collected the required skills for data scientists as described previously in Section \ref{skill-collection}.

To validate the proposed recommender system, we invited five university instructors with at least 10 years of teaching and 13 years of industrial experience, and 10 PhD students with a minimum teaching experience of one year and a minimum industrial experience of three years for a semi-structured interview\footnote{Detailed profiles of our interview participants are available on: \url{https://github.com/rezatavakoli/ICALT2020_recommender}}.

Participants gave feedback on our prototype with regards to its general objectives, logic, and potential contribution to individual learning. Each participant had to complete the following protocol:
\begin{enumerate}
    \item Learning about the research problems and the proposed approach -  15 minutes
    \item Work with our prototype dashboard - 15 minutes
    \item Going through a semi-structured interview with the help of a qualitative questionnaire\footnote{The questionnaire is available on: \url{https://github.com/rezatavakoli/ICALT2020_recommender}} - 30 minutes
\end{enumerate}

Participants received more than 250 video recommendations while working with our prototype dashboard (Each participant received 15-17 recommendations).  82.8\% of these recommendations were signalled as useful and relevant to participants' skill levels and properties. 2.8\% of the recommended videos were recognised as irrelevant, and in 14.4\% of the cases participants decided to change the recommended video. The outputs of the interviews are summarised under the three following sections.

\subsection{Objectives}
Interviewees confirmed that there is a potential value in recommending open educational videos based on labour market information. Both instructors and PhD students expressed that although there are several open educational videos on the Internet, finding the most suitable content to learners' preferences is a complicated and time-consuming task. For instance, \emph{Instructor\_2}, \emph{Instructor\_5}, \emph{Student\_10} told that personalisation of open educational content recommendation is one of the most important features of our proposed approach. Moreover, \emph{Instructor\_3} suggested that we should recognize the level of expertise that the learner needs to achieve in order to prevent over qualifying and wasting time.
\subsection{Logic}
Participants emphasized that our recommendation model can help learners in finding the most relevant videos covering particular skill areas. \emph{Student\_6} suggested that the system needs to take into account the job area of skills, which may result in fine-grained recommendations that target a specific skill for a specific job. For instance, the skill \emph{data visualization} might have different content depending on the job areas such as E-commerce or Bioinformatics. Regarding the recommendation logic, participants thought that suggesting videos based on learners' previous ratings is a novel idea and \emph{Instructor\_2} told that the system should use more user properties such as language preferences, province of residency, etc. and also properties of similar users in order to generate better recommendations.

\subsection{Contribution to Learning}
Participants confirmed that engaging with learners based on their preferences may result in better retention rates for learners. \emph{Instructor\_4} and \emph{Student\_1} valued that setting specific goals and recommending videos to learners accordingly help them focus on their skill targets and could potentially improve their learning performance. Also, \emph{Student\_4} and \emph{Student\_7} recommended to build a list of topics, which should be associated to skills, and use these topics to improve skills assessments at the beginning (setting initial expertise levels) and also during the learning process (e.g. evaluating knowledge gains after watching videos).

\section{Conclusion and Future Work}\label{conclusion}
In this paper, we demonstrated a recommender system prototype, which capitalizes on open educational videos, and built a personalized learning environment, where users can select skills and master them based on labour market information. We showed that our skill extraction approach can detect skills in vacancies with F1 score of 88.7\%. The recommender was validated with semi-structured interviews with subject matter experts. The initial results showed that participants were satisfied with 82.8\% of the generated recommendations. The validation also revealed that our proposed recommender system has high potential to support learners in constructing individual learning scenarios and direct their learning towards their individual skill targets. 

As future work, we consider to progress towards the following improvements: 1) increase user satisfaction by adding more contextual features, like user traits or more fine grained topic classification into recommendations for better personalisation, 2) incorporate more video and OER repositories and collect/predict more properties for them such as level and completeness, 3) improve self assessment and learning pathway recommendation with generating target topics for skills and finally 4) use experimental designs to further validate the system in a number of use cases with large number of learners.

\section*{Acknowledgement}
Part of this work is financially supported by the Leibniz Association, Germany (Leibniz Competition 2018, funding line “Collaborative Excellence”, Project SALIENT [K68/2017]).

\bibliographystyle{IEEEtran}
\bibliography{icalt20}

\begin{thebibliography}{10}
\providecommand{\url}[1]{#1}
\csname url@samestyle\endcsname
\providecommand{\newblock}{\relax}
\providecommand{\bibinfo}[2]{#2}
\providecommand{\BIBentrySTDinterwordspacing}{\spaceskip=0pt\relax}
\providecommand{\BIBentryALTinterwordstretchfactor}{4}
\providecommand{\BIBentryALTinterwordspacing}{\spaceskip=\fontdimen2\font plus
\BIBentryALTinterwordstretchfactor\fontdimen3\font minus
  \fontdimen4\font\relax}
\providecommand{\BIBforeignlanguage}[2]{{%
\expandafter\ifx\csname l@#1\endcsname\relax
\typeout{** WARNING: IEEEtran.bst: No hyphenation pattern has been}%
\typeout{** loaded for the language `#1'. Using the pattern for}%
\typeout{** the default language instead.}%
\else
\language=\csname l@#1\endcsname
\fi
#2}}
\providecommand{\BIBdecl}{\relax}
\BIBdecl

\bibitem{colombo2018applying}
E.~Colombo, F.~Mercorio, and M.~Mezzanzanica, ``Applying machine learning tools
  on web vacancies for labour market and skill analysis,'' 2018.

\bibitem{wowczko2015skills}
I.~Wowczko, ``Skills and vacancy analysis with data mining techniques,'' in
  \emph{Informatics}, vol.~2, no.~4.\hskip 1em plus 0.5em minus 0.4em\relax
  Multidisciplinary Digital Publishing Institute, 2015, pp. 31--49.

\bibitem{mcgill2009defining}
M.~M. McGill, ``Defining the expectation gap: a comparison of industry needs
  and existing game development curriculum,'' in \emph{Proceedings of the 4th
  International Conference on Foundations of Digital Games}.\hskip 1em plus
  0.5em minus 0.4em\relax ACM, 2009, pp. 129--136.

\bibitem{khobreh2015ontology}
M.~Khobreh, F.~Ansari, M.~Fathi, R.~Vas, S.~T. Mol, H.~A. Berkers, and
  K.~Varga, ``An ontology-based approach for the semantic representation of job
  knowledge,'' \emph{IEEE Transactions on Emerging Topics in Computing},
  vol.~4, no.~3, pp. 462--473, 2015.

\bibitem{djumalieva2018open}
J.~Djumalieva and C.~Sleeman, ``An open and data-driven taxonomy of skills
  extracted from online job adverts,'' \emph{Developing Skills in a Changing
  World of Work: Concepts, Measurement and Data Applied in Regional and Local
  Labour Market Monitoring Across Europe}, p. 425, 2018.

\bibitem{kanwar2018global}
A.~Kanwar and S.~Mishra, ``Global trends in oer: What is the future?'' 2018.

\bibitem{ha2011novel}
K.-H. Ha, K.~Niemann, U.~Schwertel, P.~Holtkamp, H.~Pirkkalainen, D.~Boerner,
  M.~Kalz, V.~Pitsilis, A.~Vidalis, D.~Pappa \emph{et~al.}, ``A novel approach
  towards skill-based search and services of open educational resources,'' in
  \emph{Research Conference on Metadata and Semantic Research}.\hskip 1em plus
  0.5em minus 0.4em\relax Springer, 2011, pp. 312--323.

\bibitem{sun2018heuristic}
G.~Sun, T.~Cui, D.~Xu, J.~Shen, and S.~Chen, ``A heuristic approach for
  new-item cold start problem in recommendation of micro open education
  resources,'' in \emph{International conference on intelligent tutoring
  systems}.\hskip 1em plus 0.5em minus 0.4em\relax Springer, 2018, pp.
  212--222.

\bibitem{ruiz2014semantically}
A.~Ruiz-Iniesta, G.~Jimenez-Diaz, and M.~Gomez-Albarran, ``A semantically
  enriched context-aware oer recommendation strategy and its application to a
  computer science oer repository,'' \emph{IEEE Transactions on Education},
  vol.~57, no.~4, pp. 255--260, 2014.

\bibitem{chicaiza2015user}
J.~Chicaiza, N.~Piedra, J.~Lopez-Vargas, and E.~Tovar-Caro, ``A user profile
  definition in context of recommendation of open educational resources. an
  approach based on linked open vocabularies,'' in \emph{IEEE Frontiers in
  Education Conference}.\hskip 1em plus 0.5em minus 0.4em\relax IEEE, 2015, pp.
  1--7.

\bibitem{chicaiza2017recommendation}
------, ``Recommendation of open educational resources. an approach based on
  linked open data,'' in \emph{Global Engineering Education Conference}.\hskip
  1em plus 0.5em minus 0.4em\relax IEEE, 2017, pp. 1316--1321.

\bibitem{wan2018learning}
S.~Wan and Z.~Niu, ``An e-learning recommendation approach based on the
  self-organization of learning resource,'' \emph{Knowledge-Based Systems},
  vol. 160, pp. 71--87, 2018.

\bibitem{sun2017towards}
G.~Sun, T.~Cui, G.~Beydoun, S.~Chen, F.~Dong, D.~Xu, and J.~Shen, ``Towards
  massive data and sparse data in adaptive micro open educational resource
  recommendation: a study on semantic knowledge base construction and cold
  start problem,'' \emph{Sustainability}, vol.~9, no.~6, p. 898, 2017.

\bibitem{lopez2014recommendation}
J.~Lopez-Vargas, N.~Piedra, J.~Chicaiza, and E.~Tovar, ``Recommendation of oers
  shared in social media based-on social networks analysis approach,'' in
  \emph{IEEE Frontiers in Education Conference}.\hskip 1em plus 0.5em minus
  0.4em\relax IEEE, 2014, pp. 1--8.

\bibitem{duffin2007oer}
J.~Duffin, B.~Muramatsu, and S.~Henson~Johnson, ``Oer recommender: A
  recommendation system for open educational resources and the national science
  digital library,'' \emph{White paper funded by the Andrew W. Mellon
  Foundation for the Folksemantic. org project}, 2007.

\bibitem{hoppe2018}
A.~Hoppe, P.~Holtz, Y.~Kammerer, R.~Yu, S.~Dietze, and R.~Ewerth, ``Current
  challenges for studying search as learning processes,'' in \emph{Linked
  Learning Workshop - Learning and Education with Web Data (LILE), in
  conjunction with ACM Conference on Web Science}, 2018.

\bibitem{mayer2005cognitive}
R.~E. Mayer, ``Cognitive theory of multimedia learning,'' \emph{The Cambridge
  handbook of multimedia learning}, vol.~41, pp. 31--48, 2005.

\bibitem{genuchten2012}
E.~van Genuchten, K.~Scheiter, and A.~Sch{\"{u}}ler, ``Examining learning from
  text and pictures for different task types: Does the multimedia effect differ
  for conceptual, causal, and procedural tasks?'' \emph{Computers in Human
  Behavior}, vol.~28, no.~6, pp. 2209--2218, 2012.

\bibitem{cooper2018moocex}
M.~Cooper, J.~Zhao, C.~Bhatt, and D.~A. Shamma, ``Moocex: Exploring educational
  video via recommendation,'' in \emph{Proceedings of the 2018 ACM on
  International Conference on Multimedia Retrieval}.\hskip 1em plus 0.5em minus
  0.4em\relax ACM, 2018, pp. 521--524.

\bibitem{Zhu2013VideoTopicCV}
Q.~Zhu, M.-L. Shyu, and H.~Wang, ``Videotopic: Content-based video
  recommendation using a topic model,'' \emph{2013 IEEE International Symposium
  on Multimedia}, pp. 219--222, 2013.

\bibitem{joulin2016bag}
A.~Joulin, E.~Grave, P.~Bojanowski, and T.~Mikolov, ``Bag of tricks for
  efficient text classification,'' \emph{arXiv preprint arXiv:1607.01759},
  2016.

\bibitem{pennington2014glove}
\BIBentryALTinterwordspacing
J.~Pennington, R.~Socher, and C.~D. Manning, ``Glove: Global vectors for word
  representation,'' in \emph{Empirical Methods in Natural Language Processing
  (EMNLP)}, 2014, pp. 1532--1543. [Online]. Available:
  \url{http://www.aclweb.org/anthology/D14-1162}
\BIBentrySTDinterwordspacing

\end{thebibliography}

\end{document}